# ANALYZING NETWORK PERFORMANCE PARAMETERS USING WIRESHARK

Ruchi Tuli

Department of Computer & Information Technology, Jubail Industrial College, Jubail Industrial City, Kingdom of Saudi Arabia

## ABSTRACT

*Network performance can be a prime concern for network administrators. The performance of the network depends on many factors. Some of the issues faced in the network performance are - Slow Internet, Bottlenecks, Loss of packets and/or retransmissions, and Excessive bandwidth consumption. For troubleshooting a network, an in-depth understanding of network protocols is required. The main objective of this research is to analyze the performance and various other parameters related to the integrity of a network in a home-based network environment using Wireshark. Network traffic is captured for different devices. The captured traffic is then analysed using Wireshark's basic statistical tools and advanced tools for various performance parameters.*

## 1. INTRODUCTION

Now days the usage of the internet has increased everywhere. The use of the internet has become a necessity for everyone. While using the network there can be many network issues such as network abuse, malicious connections, and slowing down of the internet. So network traffic monitoring becomes important to make sure that the network is running smoothly and efficiently. To monitor and analyze data traffic, a packet sniffer tool is used. Wireshark is a network packet analyzer. Packet sniffing is a technique for monitoring every packet that is sent over the network. A packet analyzer (also known as a network analyzer, protocol analyzer, or packet sniffer is a computer program that can intercept and log traffic that passes over a digital network or part of a network and then presents captured packet data in as much detail as possible [1]. Packet sniffing is usually done by hackers or malicious intruders to carry out prohibited actions such as stealing passwords, and retrieving other important data [2]. Packet sniffers are of two types: active sniffers and passive sniffers. Active sniffers can send data in the network and can be detected by other systems through different techniques whereas passive sniffers only collect data, but cannot be detected. Wireshark is a passive network sniffer [3]. Moreover, the structure of a packet sniffer consists of two parts: the packet analyzer which works on the application layer protocol, and the packet capture (pcap) which captures packets from all other layers [4]. The way packet sniffing works is divided into three processes, namely collecting, conversion, analysis, and data theft [5].

Packet sniffers can be legitimately used by a network or system administrator to monitor and troubleshoot network traffic. Using the information captured by the packet sniffer an administrator can identify erroneous packets and use the data to pinpoint bottlenecks and help maintain efficient network data transmission. The security threat presented by sniffers is their ability to capture all incoming and outgoing traffic, including clear text passwords and usernames or other sensitive material [6].

## 1.1. Working of a Packet Sniffer

A packet passes through many intermediate devices while travelling from source to destination. Each device in the network has NIC's physical address which is uniquely identified. When the device is sending the packet, the data is received by all the devices in the network. Any node whose NIC is set in the promiscuous mode receives all information that travels in a network. When the NIC is set on promiscuous mode, the machine can see all traffic on the segment. However, when the NIC puts in promiscuous mode for one machine, the NIC takes and gathers all frames and packets on the network even if that frames and packets do not destine for that machine, which in this situation is called sniffer. The sniffer begins reading all information entered into the machine via NIC [6] [7].

## 1.2. Wireshark – Network Packet Analyzer

It is invented by the scientist Gerald Combs in late 1997 for trucking and recognizing the network's problems and monitoring the data traffic. He named it Ethereal until May, 2006, and after that its name changed to Wireshark. It is an open-source software, free and GUI packet analyzer tool that is written in C programming language and released under GNU General Public License (GPL). It runs on a variety of Unix-like operating systems including Mac OS X, Linux, Solaris platform, as well as the Microsoft Windows operating system. Command Line Interface (CLI) of Wireshark is called TShark to enable the user to deal with it via commands.

Wireshark is designed for capturing packets from live networks and also browsing previously saved captured data files. The supported format of packet capture is the "PCAP" file format. It displays the captured data in byte and hexadecimal formats showing different types of used packets and protocols. It also allows the user to assemble the packet data into a TCP stream [7].

The important point to be noted here is that to run the Wireshark tool, it sets the NIC to a promiscuous mode for enabling the sniffer to see all traffic on that interface, not just the traffic that is addressed to one of the configured interfaces. Besides the promiscuous mode, it may enable the port mirroring to any points of the network when the promiscuous mode does not cover all networks [8] [9].

Nevertheless, in this study, the sniffing packet was only used for monitoring and analyzing network traffic.

## 2. RELATED WORK

This section discusses the related work of various authors. Henry & Agana [3] monitored the Intranet traffic by sniffing the packets at the Local Area Network (LAN) server end to provide security and control. They analyzed the IP and MAC address sources and destinations of the frames, Ethernet, UDP, and HTTP protocols and concluded that these can assist network administrators to make informed decisions on possible threats that the network can be exposed to.

Ashaari, Kassim et.al [10] analyzed the high-speed fiber internet connections called TM UniFi in Malaysia with the impact of multiple device connections or users' services. They used the Wireshark network packet analyzer to analyze the QoS on its traffic, packet transfer, RTT, latency, and throughput.

Siswanto et. al [2] analyzed the traffic in a vocational school to find out the number of users who access the internet and use bandwidth. They concluded that the Wireshark tool is very helpful as it has supportive features to analyze networks.

P. Chaudhary & A.K. Singh [11] analyzed TCP & UDP packets for communication purposes over the internet using Wireshark. They concluded that Wireshark identifies every packet in detail. It helps troubleshoot the network for example it can identify denial of service attacks.

Petar Boyanov [12] analyzed the performance and connection analysis in a computer network. They used Wireshark for analyzing the networks in a university.

## 3. METHODOLOGY

The system was designed using 6 systems, consisting of smart TV, PS5, 2 laptops, and 2 mobile phones connected to a common access point. The access point was assigned a static IP address 192.168.100.1. Wireshark was deployed on one of the laptops (sniffer laptop) with IP address 192.168.100.15 and traffic was captured. The laptop is connected to the router via Ethernet as shown in Figure 1. The NIC of the sniffer laptop was set to promiscuous mode and was running the Wireshark program, thus capturing live packets in the network. A total number of 3,42,323 packets were captured and analyzed. One of the greatest strengths of Wireshark is its various statistical tools. Some basic, as well as advanced statistical tools were used to study the network performance.

### 3.1. System Design

Figure 1 shows the system design. The System consists of 6 devices. The devices are connected to a common access point.

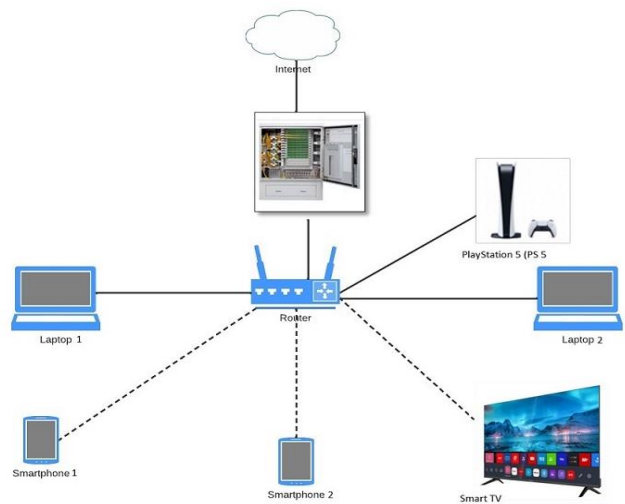

Figure 1: Network Topology

Figure. 1 above shows the network design for the experiment. The IP addresses of the devices attached are shown in the table.

Table 1: IP addresses of the devices

| Device | IP address |
|---|---|
| Router | 192.168.100.1 |
| Laptop 1 | 192.168.100.15 |
| Laptop 2 | 192.168.100.47 |
| Smartphone 1 | 192.168.100.62 |
| Smartphone 2 | 192.168.100.23 |
| Smart TV | 192.168.100.35 |
| PlayStation 5 | 192.168.100.47 |

## 3.2. Experimental Analysis

In this section, first, we will discuss some basic statistical tools then followed by advanced statistical tools. The advanced statistical tools discussed are – I/O graphs, Steven's window, and TCP trace window. Figure 2(a) below shows the interface overview of Wireshark capture and Figure 2(b) shows the total number of packets captured.

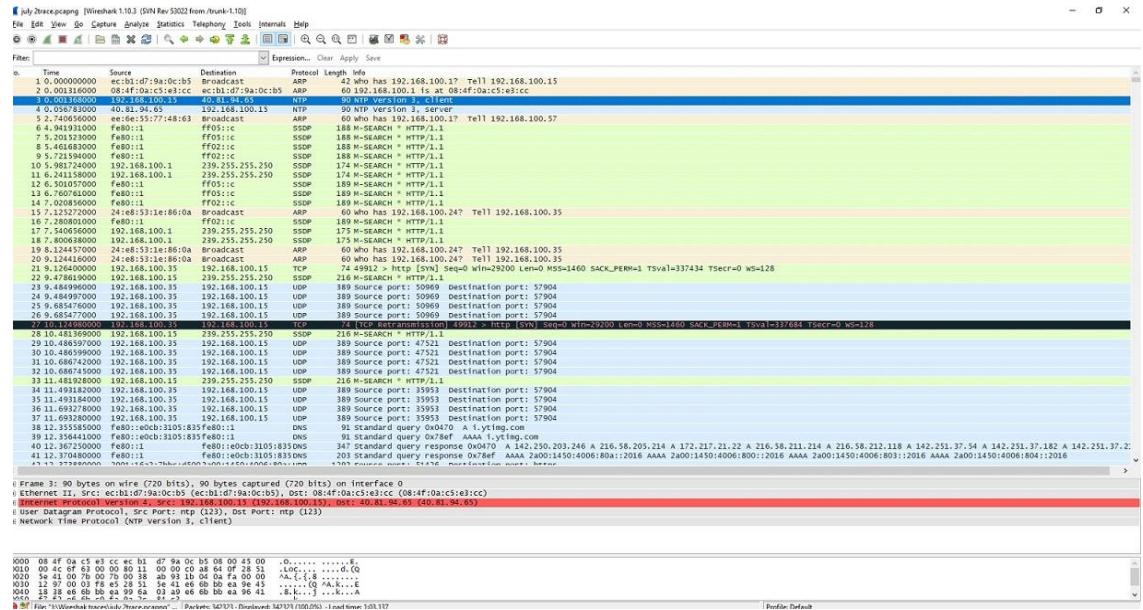

Figure 2 (a): Wireshark Interface Overview

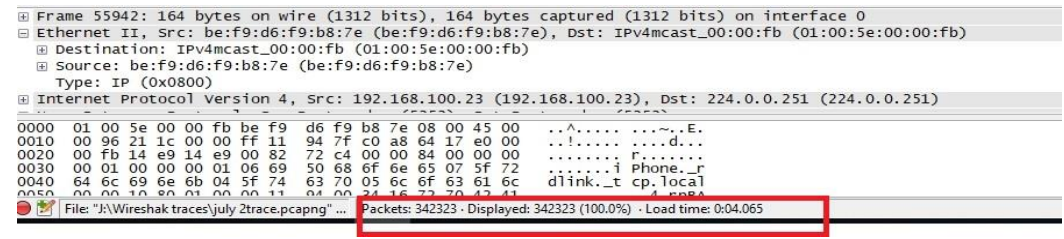

Figure 2 (b): Wireshark Interface Overview

## 3.3. Basic Statistical Tools

The basic statistical tools that are discussed to monitor the network performance are – Summary, resolved addresses, protocol hierarchy, and conversations.

### 3.3.1. Summary

Fig 3 shows the general information of the captured file.

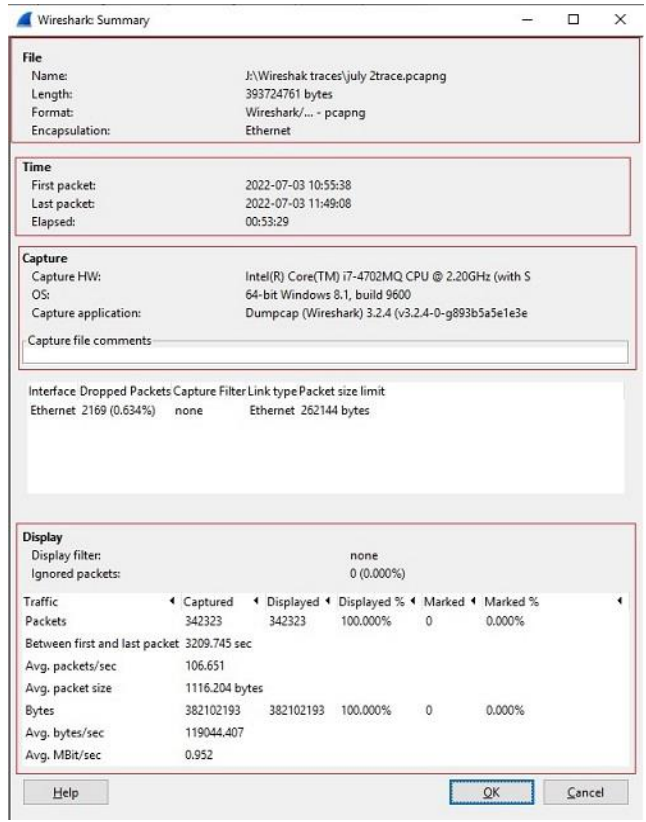

Figure 3: Wireshark Summary Window

### 3.3.2. Resolved Addresses

This option resolves DNS translations of the captured IP addresses [13] [14]. Figure 4 shows the resolved DNS addresses.

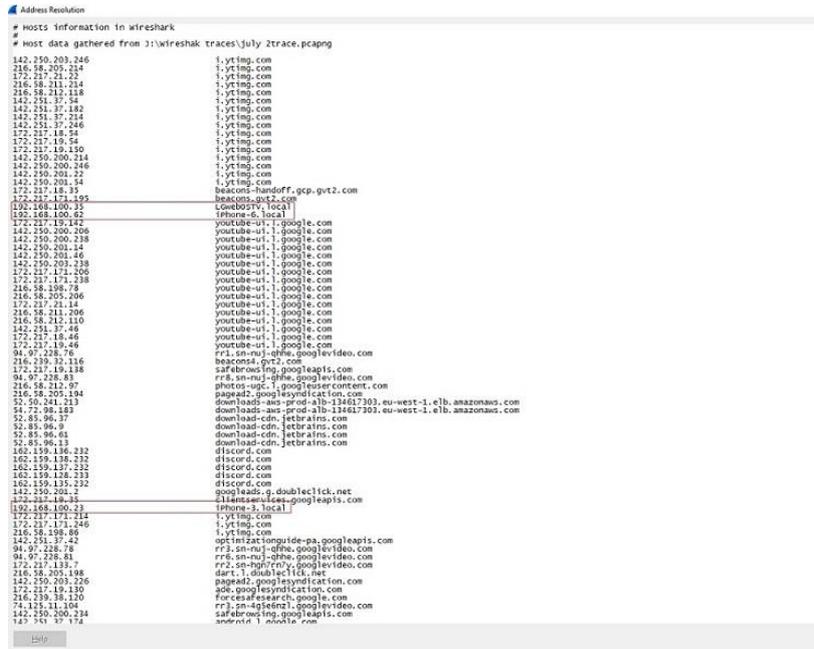

Figure 4: Resolved Addresses

### 3.3.3. Protocol Hierarchy

This option gives us data about the protocol distribution in the captured file conversations [13]. The Protocol Hierarchy window provides us with an overview regarding the distribution of protocols used in the communication process and how a network administrator can identify unusual activities [15]. Figure 5 shows the protocol hierarchy.

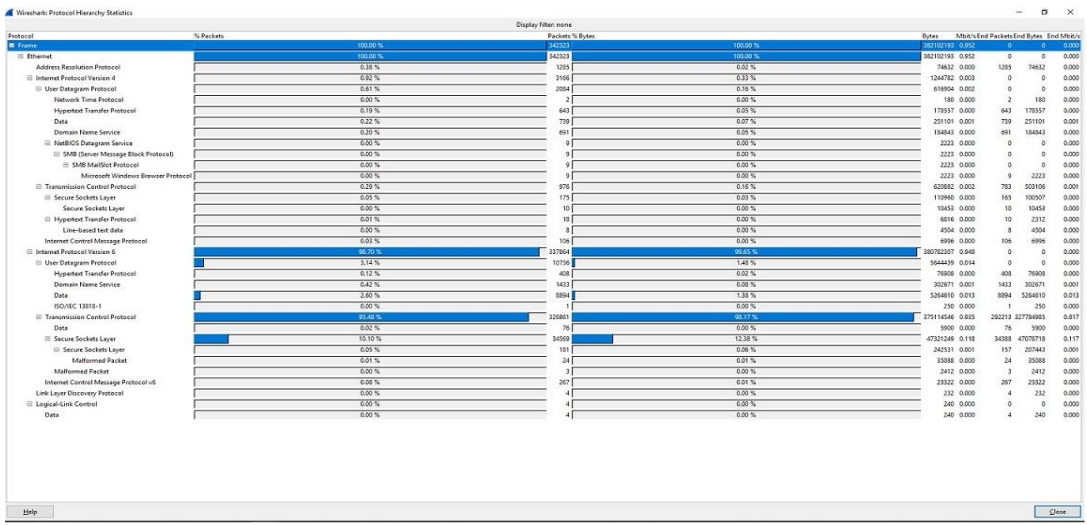

Figure 5: Protocol Hierarchy

### 3.3.4. Conversations

When two devices are connected to the network, they are supposed to communicate; this is considered normal behaviour [15]. However, when thousands of devices are connected to the network and we want to figure out the most active device that is generating too much traffic, then in that instance, the Conversations window will be quite useful. The conversation window gives the option to choose between layer 2 Ethernet statistics, layer 3 IP statistics, or layer 4 TCP or UDP statistics. Figure 6 shows the conversation window.

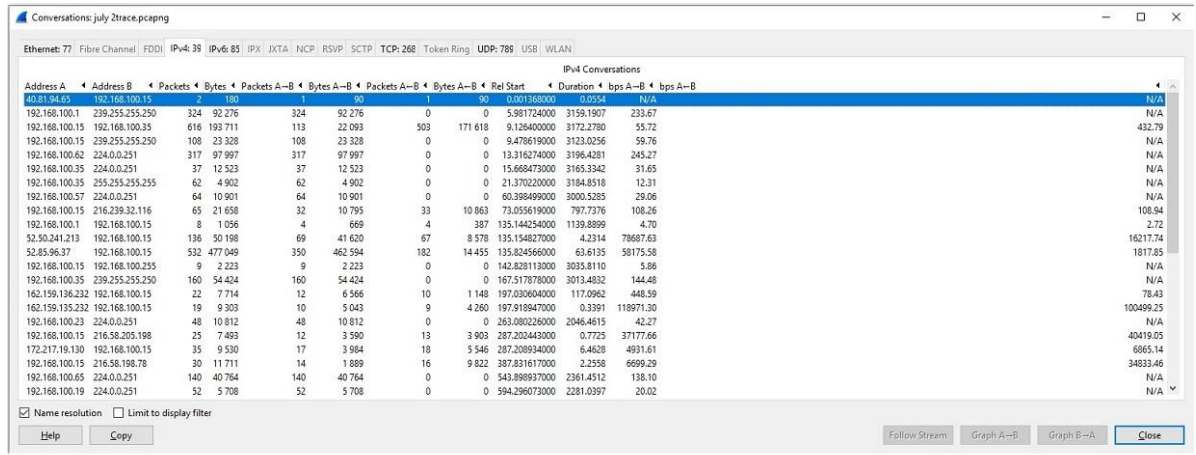

Figure 6: Conversation window

As shown in the figure above conversation window show addresses, packet counters, byte, the start time of the conversation ("Rel Start") or ("Abs Start"), the duration of the conversation in seconds, and the average bits (not bytes) per second in each direction.

## 3.4. Advanced Statistical Tools

The advanced statistical tools discussed are –various graphs and Expert Info.

### 3.4.1. Graphs

Wireshark offers graphing capabilities which can present captured packets in an interesting format that makes the analysis process much more effective and easy to adapt [14] [15]. There are multiple types of graphs available.

#### 3.4.1.1. I/O Graph

I/O graphs tell us the basic status of a network, and let us create filters. Matching network performances and differentiating a specific protocol becomes easy [14] [15]. The figure below shows the graph.

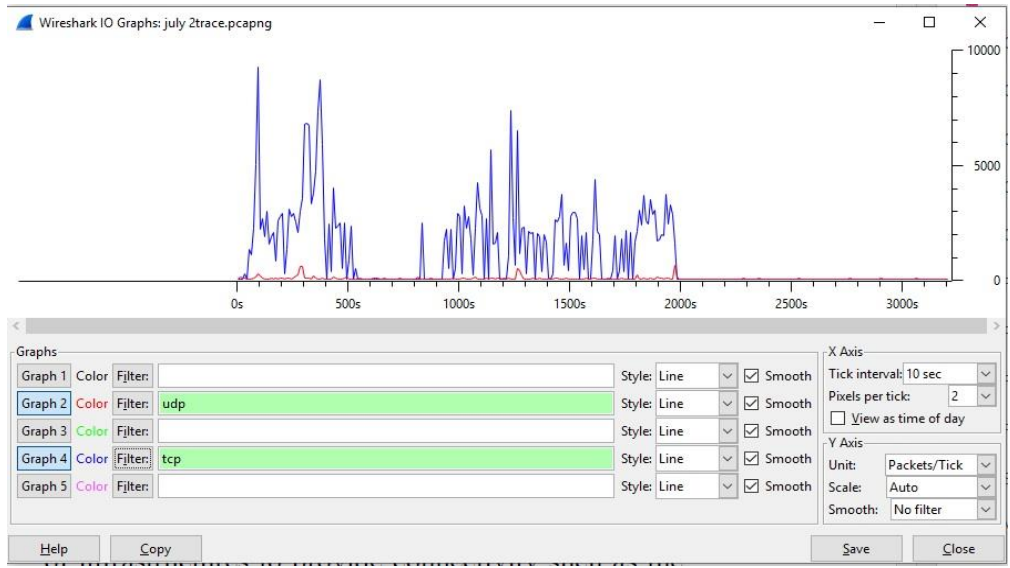

Figure 7: I/O Graph

In the figure above we applied the filter UDP and TCP to observe the traffic related to these protocols. We can see highs and lows in the network traffic, which can be used to rectify problems or can even be used for monitoring purposes. In the preceding graph, the data on the x-axis represents the time in seconds and the data on the y-axis represents the number of packets per tick [15]. The scale for the x and y-axis can be altered if needed, where the x-axis will have a range between 10 and 0.001 seconds, and the y-axis values will range between packets/bytes/bits.

The graph above displays the packet captured for TCP & UDP protocols. From the graph, we can analyze that the maximum number of 700 UDP packets are captured between the time interval 1900s to 2000s. The highest number of TCP packets captured is 9000 at a time interval between 0s to 100s.

### 3.4.1.2. Flow Graph

This feature in Wireshark provides troubleshooting capabilities in scenarios like facing a lot of dropped connections, lost frames, retransmission traffic, and more. Flow graphs enable us to create a column-based graph, which summarizes the flow of traffic between two endpoints [15].

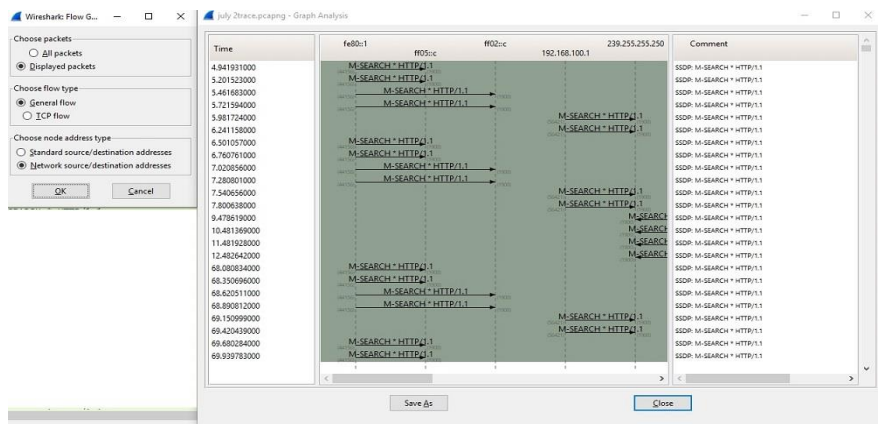

Figure 8: Flow Graph

### 3.4.1.3. Round trip time (RTT) Graph

This graph is used by network admins to identify any congestion or latency that can make your network perform slowly [15].

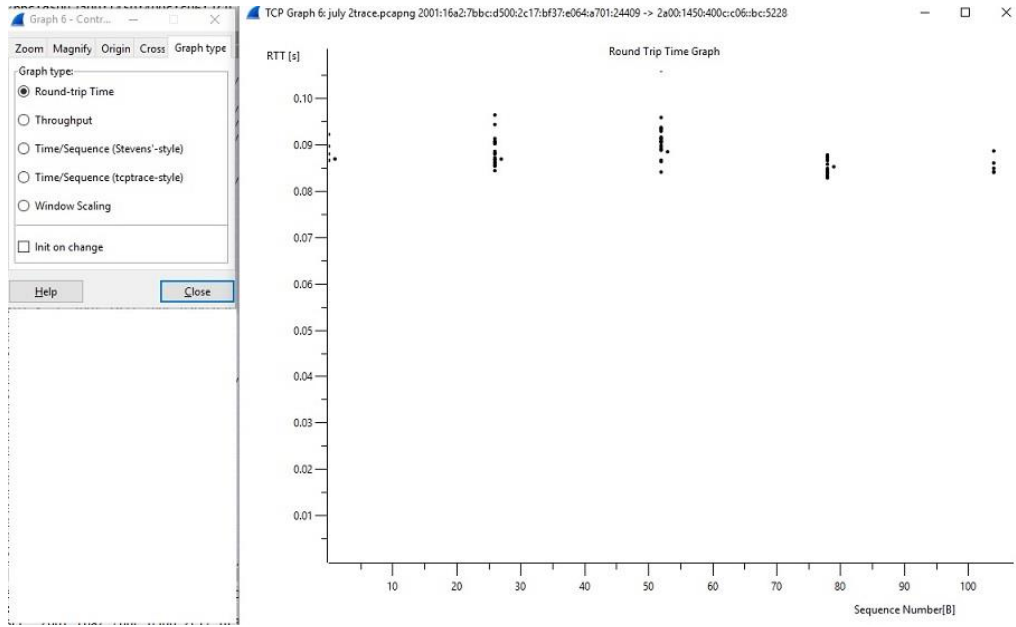

Figure 9: RTT Graph

In the above graph, at some points latency can be observed.

### 3.4.1.4. Throughput Graph

Throughput graph is similar to the I/O graph and is used for depicting the traffic flow.

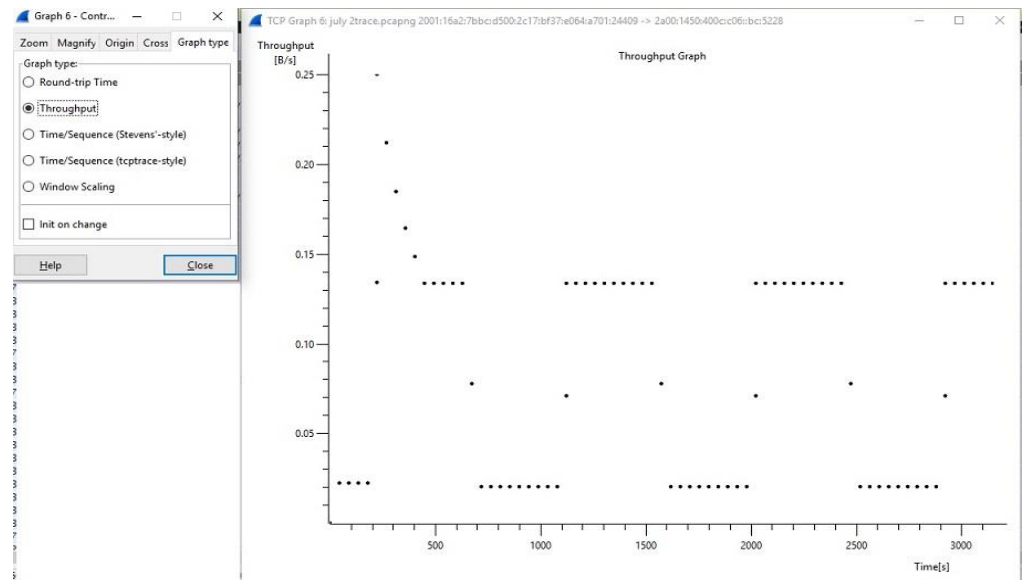

Figure 10: Throughput Graph

**3.4.1.5.  Time-sequence graph (tcptrace)**

This graph shows the TCP stream over time. The traffic depicted in this graph is unidirectional. Time-sequence graph gives us an idea about the segments that are currently traveling, the acknowledgments for segments that we've received, and the buffer area that the client is capable to hold.

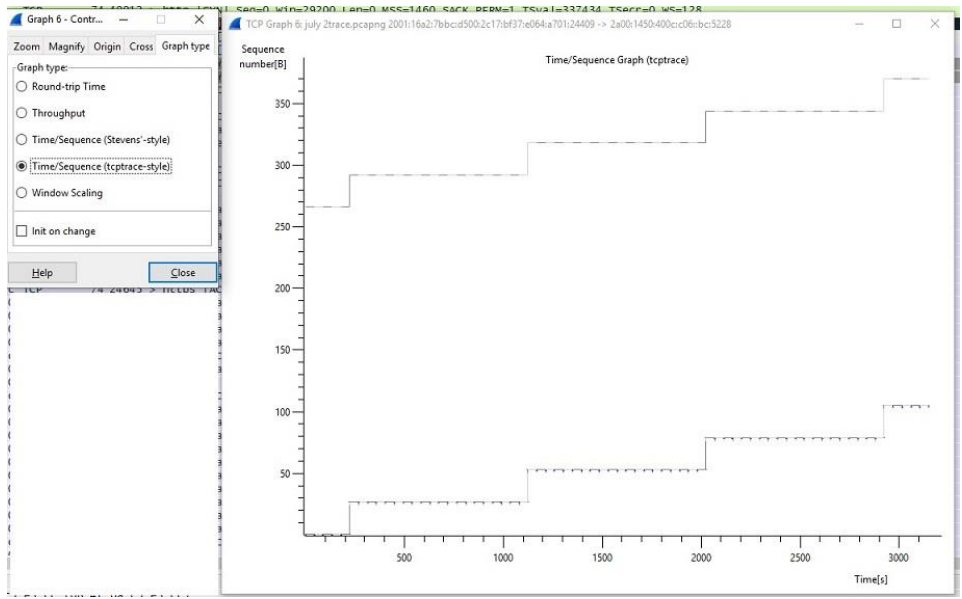

Figure 11: tcptrace Graph

**3.4.2.  Expert Info**

The information in the Expert Info's dialog is populated by the dissectors that enable the translation of every protocol that is well known to Wireshark. The purpose of the Expert Info dialog is to keep the user aware of the specific states that users should know about [15]. Figure 12 shows the expert info.

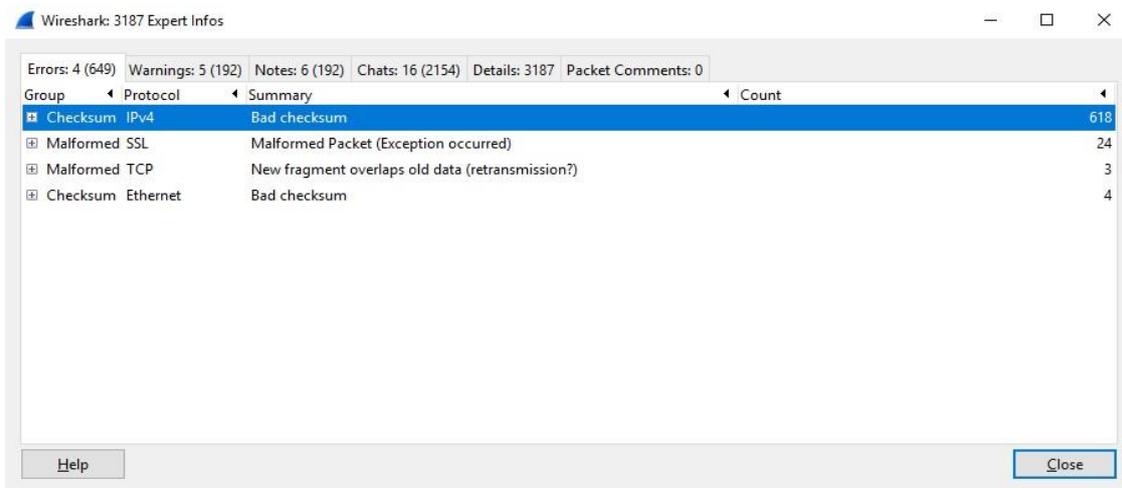

Figure 12: Expert Info

The window has 6 tabs – Errors, warnings, Notes, chats, details and packet comments. The main objective of the Expert Info dialog is to find the errors/anomalies present in a trace file. Finding the network problems in the trace file for a novice user becomes a lot easier and faster. Viewing the Expert Info dialog can give a better idea about the unusual behaviour of network packets [15].

# 4. RESULT AND DISCUSSION

The graphs depicted in the previous section show that Wireshark offers many capabilities to analyze and summarize various connections in the networks.

Wireshark Interface view updates the packet list pane in real time during capture. As shown in Fig 2(a), time of capturing the packet, Source & destination IP, protocol name, length of the packet (in bytes) and Info about the packet is shown on the interface screen. Total number of packets captured and dropped is also shown as depicted in figure 2(b).

The Summary option is helpful for getting information such as the file name, the total number of bytes captured, the start time, end time, and duration of the capture, as well as hardware details about the PC that was used to record the packets, interface details, and general statistics about the capture.

Resolved Addresses option is useful to know the host name of the IP addresses captured.
The protocol hierarchy window displays the protocol tree for all protocols in the capture. Each row contains one protocol's statistical values. Network administrators can use the protocol hierarchy option to detect any unusual activity in the network.

The conversation window includes tabs for layer 2 Ethernet statistics, layer 3 IP statistics, and layer 4 TCP or UDP statistics. This can be used to locate and isolate broadcast storms on Layer 2 (Ethernet), as well as to connect in parallel to the internet router port and determine who is loading the line to the ISP (Layer 3/Layer 4 (TCP/IP)). If a network administrator notices a lot of traffic going out to port 80 (HTTP) on a specific IP address on the internet, simply copying the address into a browser can provide information about the most popular website among users [13].

The Wireshark I/O graph feature provides an overview of the overall traffic seen in a capture file, which is typically measured in bytes or packets per second (the user can always change to bits/bytes per second). The x-axis is the tick interval per second by default, and the y-axis is the packets per tick (per second)[13].

Flow graph is one of Wireshark's most useful features, assisting network administrators with troubleshooting capabilities in scenarios such as dealing with a large number of dropped connections, lost frames, retransmission traffic, and more. Flow graphs allow us to create a column-based graph that summarizes the flow of traffic between two endpoints, and the results can even be exported in a simple text-based format. This is the simplest method of verifying the client-server connection.

The round-trip time (RTT) is the time it takes to receive the acknowledgment (ACK) for a sent packet; that is, for every packet sent from a host, an ACK is received (TCP communication), which determines the packet's successful delivery. The total time spent from packet transfer to ACK is referred to as round trip time [15]. The RTT graph allows us to examine the round trip between sequence numbers and the time they were acknowledged [13].

The throughput graph is similar to the IO graph, which depicts traffic flow [15]. However, there is one significant difference: throughput graphs depict unidirectional traffic, whereas IO graphs depict traffic in both directions [13]. The Throughput graph can differ for each TCP packet selected in the list pane. The throughput graph allows network administrators to examine a connection's throughput. Depending on the application, network administrators can use this graph to check for instabilities.

For years, analysts have used the tcptrace graph to visualize the efficiency of TCP data transfers. It allows us to track the increase in sequence number over time, the received TCP window, bytes in flight, retransmissions, and acknowledged data.

One of Wireshark's most powerful features is its ability to analyze network phenomena and suggest a possible cause. It provides detailed information on network performance and problems, in conjunction with other tools. The expert window is extremely useful in providing a list of Wireshark-discovered events and network problems.

# 5. CONCLUSION

Wireshark, as described above, is extremely useful for analyzing and monitoring a network, as well as detecting various threats that can slow down a network. Wireshark is a useful tool that provides a plethora of functions that aid in the analysis of various network problems, including those caused by poor network configuration or device failures, as well as a variety of external and internal LAN attacks. The first step in resolving network problems is to identify where the problem occurred in the network and what caused performance loss. Wireshark is a powerful tool that can be used not only by network administrators, IT specialists, and cyber security professionals, but also by home users to analyze unusual network behaviours and to measure network performance.

## ACKNOWLEDGEMENTS

We would like to thank everyone who has contributed to this research directly or indirectly.

## AUTHOR

**Dr, Ruchi Tuli,** is presently working as an Assistant Professor of Computer Science in the Department of Computer & Information Technology at Jubail Industrial College, Jubail, Kingdom of Saudi Arabia. She obtained her Ph.D. degree in computer science in 2011 from Singhania University. She has more than 15 years of teaching experience. She has in her credit many papers published in reputed international journals. She is also the author of 2 books – "Recovery in Mobile and ad hoc networks" and "Understanding Computer Networks.
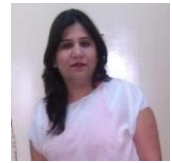